\documentclass[conference]{IEEEtran}

\usepackage{amsmath}
\usepackage{balance}
\usepackage[noadjust]{cite}

\ifCLASSINFOpdf
  \usepackage[pdftex]{graphicx}
\else
\fi
\usepackage{array}

\usepackage{multirow}

\begin{document}
\title{\Huge SOT-MRAM based Sigmoidal Neuron for Neuromorphic Architectures}
%Power-Error-Product Efficient Digital  Probabilistic Interpolation Circuits for Analog-Assisted Deep Belief Networks}


% \author{Brendan Reidy, and Ramtin Zand}
% \affiliation{Department of Computer Science and Engineering, University of South Carolina, Columbia, SC, 29208 USA}
%   \city{Orlando} 
%   \state{Florida} 
%   \postcode{32816}

\author{\IEEEauthorblockN{Brendan Reidy and Ramtin Zand}
\IEEEauthorblockA{Department of Computer Science and Engineering, University of South Carolina, Columbia, SC 29208. (ramtin@cse.sc.edu)
}
}
\maketitle

\begin{abstract}
In this paper, the intrinsic physical characteristics of spin orbit torque (SOT) magnetoresistive random-access memory (MRAM) devices are leveraged to realize sigmoidal neurons in neuromorphic architectures. Performance comparisons with the previous power- and area-efficient sigmoidal neuron circuits exhibit $74\times$ and $12\times$ reduction in power-area-product values for the proposed SOT-MRAM based neuron. To verify the functionally of the proposed neuron within larger scale designs, we have implemented a circuit realization of a $784 \times 16 \times 10$ SOT-MRAM based multiplayer perceptron (MLP) for MNIST pattern recognition application using SPICE circuit simulation tool. The results obtained exhibit that the proposed SOT-MRAM based MLP can achieve accuracies comparable to an ideal binarized MLP architecture implemented on GPU, while realizing orders of magnitude increase in processing speed.

%In this paper, a probabilistic interpolation recoder (PIR) circuit is developed for deep belief networks (DBNs) with probabilistic spin logic (p-bit)-based neurons. To verify the functionality and evaluate the performance of the PIRs, we have implemented a $784 \times 200 \times 10$ DBN circuit in SPICE for pattern recognition application using the MNIST dataset. The PIR circuits are leveraged in the last hidden layer to interpolate the probabilistic output of the neurons, which are representing different output classes, through sampling the p-bit's output and accumulating them in a defined sampling time window. The PIR circuit is proposed as an alternative for conventional interpolation methods which were based on using a resistor-capacitor circuit to integrate the neuron's output, followed by an analog-to-digital converter to generate the digital output. The circuit simulation results of PIR circuits exhibit at least 59\%, 67\%, and 24\% reductions in power, energy, and energy-error-product, respectively, compared to previous techniques, without using any of the area-consuming analog components in the interpolation circuit.    

\end{abstract}


% \begin{IEEEkeywords}
% Magnetic random access memory (MRAM), multi-layer perceptron (MLP), neuromorphic, sigmoid neuron, spin orbit torque (SOT).
% \end{IEEEkeywords}

% \fi
\IEEEpeerreviewmaketitle

\section{Introduction}

The neuromorphic computing is the concept of embodying the physical processes that underlie the computations of biological neural networks (NNs) within the physics of the very large-scale integration (VLSI) circuits, as opposed to the conventional power-hungry approaches which emulate the mathematical behavior NNs on conventional computing systems such as GPUs. Recently, various beyond CMOS technologies have been investigated to be leveraged within neuromorphic circuits and architectures, among which memristive devices are one of the most promising solutions \cite{Neuromemristive2020}.    

Memristors have been widely used within both synapse and neuron circuits, and provide significant advantages such as small on-chip area, non-volatility, and low-power dissipation \cite{Neuromemristive2020}. However, they suffer from severe reliability issues such as high device-to-device (D2D) and cycle-to-cycle (C2C) variations \cite{variation1} and low endurance \cite{Endurance1}. On the other hand, spintronic devices have shown some reliability advantages over other memristive devices. For instance, spin orbit torque (SOT) magnetoresistive random-access memory (MRAM) \cite{Liu2012} have exhibited infinite write endurance, which is a desirable feature for in-circuit training that can be used to alleviate variation challenges \cite{variation1} in neuromemristive architectures \cite{Neuromemristive2020}. While SOT-MRAM devices have been previously used within in-memory computing platforms as a hardware accelerator for artificial neural networks \cite{inmemory}, herein we will go beyond the previous work and utilize the intrinsic characteristics of SOT-MRAM cells within the neuromorphic architecture as a natural building block for both synapses and neurons.

\section{SOT-MRAM based Neurons and Synapses}
\begin{figure}
\centering
\includegraphics[scale=0.4]{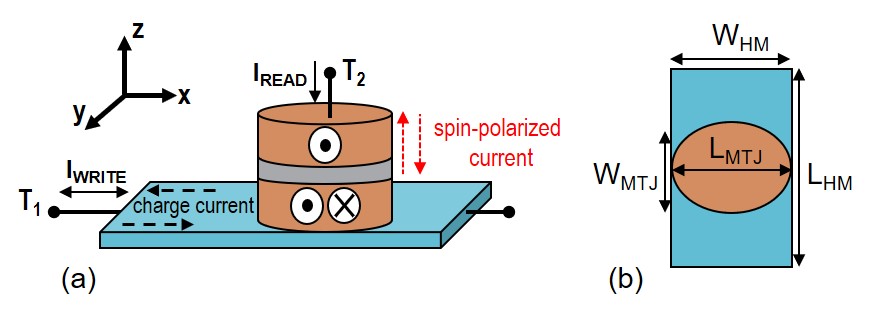}
\caption{(a) SOT-MRAM cell. Positive current along $+x$ induces a spin injection current $+z$ direction. The injected spin current produces the required spin torque for aligning the magnetic direction of the free layer in $+y$ directions, and vice versa. (b) SOT-MRAM Top view.}
\label{fig:sotmram}
\end{figure}

Fig. \ref{fig:sotmram} shows a simplified structure of a SOT-MRAM cell, which includes a magnetic tunnel junction (MTJ) with two ferromagnetic (FM) layers, which are separated by a thin oxide layer. MTJ has two different resistance levels, which are determined according to the angle ($\theta$) between the magnetization orientation of the FM layers. The resistance of the MTJ in parallel (P) and antiparallel (AP) magnetization configurations can be obtained using the following equations \cite{Zhang2012CompactModeling}:

\begin{equation} 
\label{EqR} 
\small 
\begin{aligned}
R(\theta) = & \frac{2R_{MTJ}(1 + TMR)}{2 + TMR ( 1 + \cos\theta)}\ \\= 
& \begin{cases} 
R_P=R_{MTJ}, & \theta=0  \\ 
R_{AP}=R_{MTJ}(1+TMR), & \theta=\pi 
\end{cases} 
\end{aligned}
\end{equation}

% \begin{equation} \small \label{EqTMR} TMR (T,V_b)=  \frac{2P^2(1-\alpha_{sp}T^{3/2})^2}{1-P^2(1-\alpha_{sp}T^{3/2})^2}.\frac{1}{1+(\frac{V_b}{V_0})^2}\    \end{equation} 
\begin{equation} \small \label{EqTMR} TMR(T,V_b)= \frac{TMR_0/100}{1+(\frac{V_b}{V_0})^2}\    \end{equation}

\noindent where $R_{MTJ} = \frac{RA}{Area}$, in which the resistance-area product (RA) value of the MTJ depends on the material composition of its layers. TMR is the tunneling magnetoresistance, which relies on temperature (T) and bias voltage ($V_b$). $V_0$ is a fitting parameter, and $TMR_0$ is a material-dependent constant 
%$P$ is the spin polarization factor, $V_0$ is a fitting parameter, and $\alpha_{sp}$ is a material-dependent constant.

In the MTJ structure, the magnetization direction of electrons in one of the FM layers is fixed (pinned layer), while the electrons' directions in the other FM layer (free layer) can be switched. In \cite{Liu2012}, Liu et al. have shown that passing a charge current ($I_c$) through a heavy metal (HM) generates a spin-polarized current ($I_s$) using the spin Hall Effect (SHE), which can switch the magnetization direction of the free layer, as shown in Fig. \ref{fig:sotmram}. The ratio of the generated spin current to the applied charge current is normally greater than one leading to an energy-efficient switching operation \cite{zandTVLSI}. Herein, we will use SOT-MRAM devices as a building block for both synapse and neuron circuits.

\begin{table}
\centering
\footnotesize
\caption{Parameters of the SHE-MRAM device \cite{Zhang2012CompactModeling}.}
\label{table:sheparameters}
\begin{tabular}{c c c} \hline  
{\bf Parameter} & {\bf Description} & {\bf Value} \\ \hline
\multirow{2}{*}{$MTJ_{Area}$} &  \multirow{2}{*}{$l_{MTJ}\times w_{MTJ} \times \frac{\pi}{4}$} & \multirow{2}{*}{$50nm \times 30nm \times \frac{\pi}{4}$}  \\ 
		{}&{}&{} \\ 
\multirow{2}{*}{$HM_{V}$} &  \multirow{2}{*}{$l_{HM}\times w_{HM} \times t_{HM}$} & \multirow{2}{*}{$100nm \times 50nm \times 3nm$}  \\ 
{}&{}&{} \\ 
%{$\alpha$} &    {Gilbert Damping factor} & {$0.007$}  \\ 
%{$P$}&    {Spin Polarization } & {$0.52$}  \\ 
%{$M_{S}$}&    {Saturation magnetization } & {$7.8e5 A.m^{-1}$}  \\ 
{$RA$}&    {resistance-area product} & { 10 $\Omega.\mu m^2$}  \\
{$V_{0}$}&    {Fitting parameter} & {0.65}  \\ 
{$TMR_0$}&    {tunneling magnetoresistance} & {100}  \\ \hline
%{$\alpha_{sp}$}&    {Material-dependent constant } & {2e-5 }  \\ \hline
\end{tabular}
\end{table}

\subsection{SOT-MRAM Based Neuron}
\begin{figure}[!t]
\centering
\includegraphics[scale=0.37]{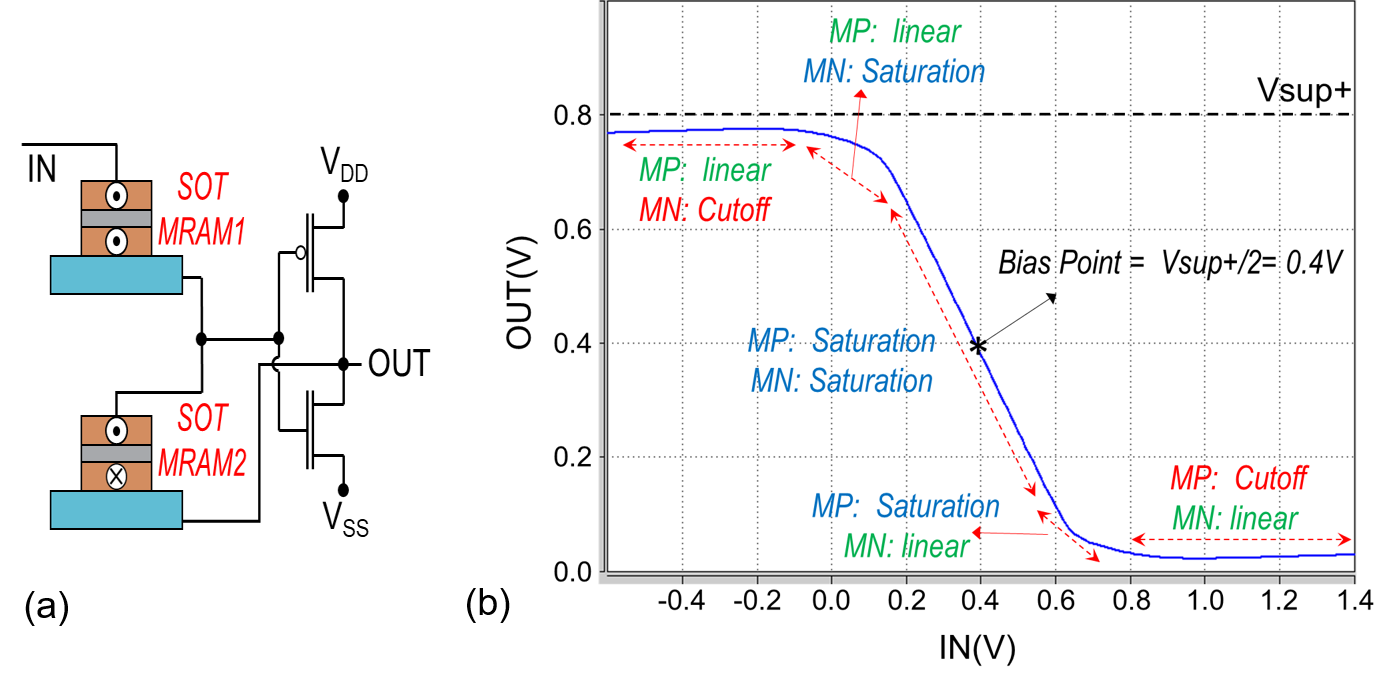}
\vspace{-2mm}
\caption{(a) The SOT-MRAM based neuron, (b) The VTC curves showing various operating regions of PMOS (MP) and NMOS (MN) transistors.} 
\label{fig:neuron}
\end{figure}

Fig. \ref{fig:neuron} (a) shows the structure of the proposed neuron, which includes two SOT-MRAM devices and a CMOS-based inverter (2T-2R). The magnetization configuration in SOT-MRAM1 is required to be in $P$ state, while SOT-MRAM2 is in $AP$ state. The SOT-MRAMs in the neuron's circuit operate as a voltage divider, which reduces the slope of the linear operating region in the CMOS inverter's voltage transfer characteristic (VTC) curve. The reduction in the slope of linear region in the CMOS inverter creates a smooth high-to-low output voltage transition, which enables the realization of the activation function behavior desirable for sigmoid neurons. 

% \begin{figure}[!t]
% \centering
% \includegraphics[scale=0.22]{Figures/sig.png}
% \caption{The VTC curves for SOT-MRAM based Sigmoid neuron showing various operating regions of PMOS (MP) and NMOS (MN) transistors.} 
% \label{fig:sigtan}
% \end{figure} 

In order to verify the functionality of our proposed neuron, first we created a Verilog-A model of the SOT-MRAM device using equations (\ref{EqR}) and (\ref{EqTMR}), and the parameters listed in Table \ref{table:sheparameters} \cite{Zhang2012CompactModeling}. Next, we utilized the developed model along with 14nm HP-FinFET PTM library to realize the circuit implementation of the neuron. Fig. \ref{fig:neuron} (b) shows the SPICE circuit simulation results of the proposed SOT-MRAM based neuron using $V_{DD}=0.8V$ and $V_{SS}=0V$ voltages, which validates the desired sigmoidal behavior for neurons.  %Moreover, the same neuron structure can be utilized to implement a $tanh$ activation function by changing the supply voltages to $V_{DD}=0.8V$ and $V_{SS}=-0.8V$, as shown in Figure \ref{fig:sigtan} (b). Both the sigmoid and $tanh$ activation functions are biased around $\frac{1}{2}(V_{DD}-V_{SS})$ voltage, i.e. 0.4(V) and 0(V) for sigmoid and $tanh$, respectively. The effect of non-zero bias voltage for the sigmoid function can be removed at both architecture- and algorithm-level, as described in next sections.    

\begin{figure}[!t]
\centering
\includegraphics[scale=0.34]{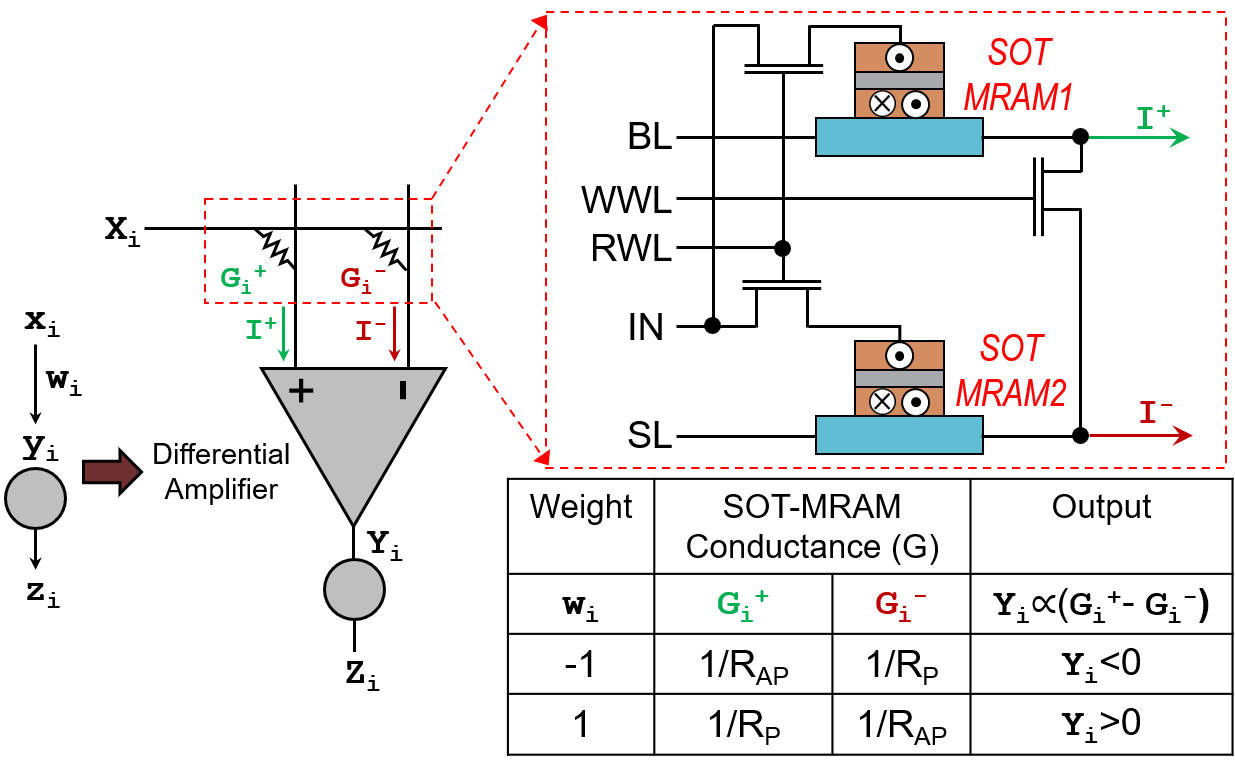}
\caption{The SOT-MRAM based binary synapse.} 
\label{fig:synapse}
\end{figure}

\subsection{SOT-MRAM Based Synapse}
SOT-MRAM cell are capable of realizing two resistive level, i.e. $R_P$ and $R_{AP}$. The combination of two SOT-MRAM cells and a differential amplifier can produce the positive and negative weights required for implementation of a binary synapse. Fig. \ref{fig:synapse} shows a neuron with $Y_i=X_i \times W_i$ as the input of the neuron, where $X_i$ is the input signal and $W_i$ is a binarized weight. The corresponding circuit implementation is also shown in the figure, which includes two SOT-MRAM cells and a differential amplifier as synapse. The output of the differential amplifier ($Y_i$) is proportional to ($I^+-I^-$), where $I^+ = X_iG_i^+$ and $I^- = X_iG_i^-$. Thus, $Y_i \propto X_i(G_i^+-G_i^-$) in which $G_i^+$ and $G_i^+$ are the conductance of SOT-MRAM1 and SOT-MRAM2, respectively, that can be tuned as shown in Fig. \ref{fig:synapse} to realize negative and positive weights in a binary synapse. For instance, for $\textbf{W}_i=-1$, SOT-MRAM1 and SOT-MRAM2 should be in $AP$ state and $P$ states, respectively. According to Eq. (\ref{EqR}) $R_{AP}>R_P$, which means $G_{AP}<G_P$ since $G=1/R$, therefore $G_i^+ < G_i^+$ and $Y_i<0$.    

%It is worth noting that similar components can be utilized to implement ternarized weights ($\textbf{W}_i \in \{-1, 0, +1\}$). However, as shown in Fig. \ref{fig:synapse} (b), one extra transistor and two extra control lines are required to realize each ternary synapse, which can impose a significant area overhead to the proposed neuromorphic MLP architecture described in next Section. 

\section{Proposed SOT-MRAM based MLP Architecture}
\begin{figure}[!t]
\centering
\includegraphics[width=3.4in,height=2.2in]{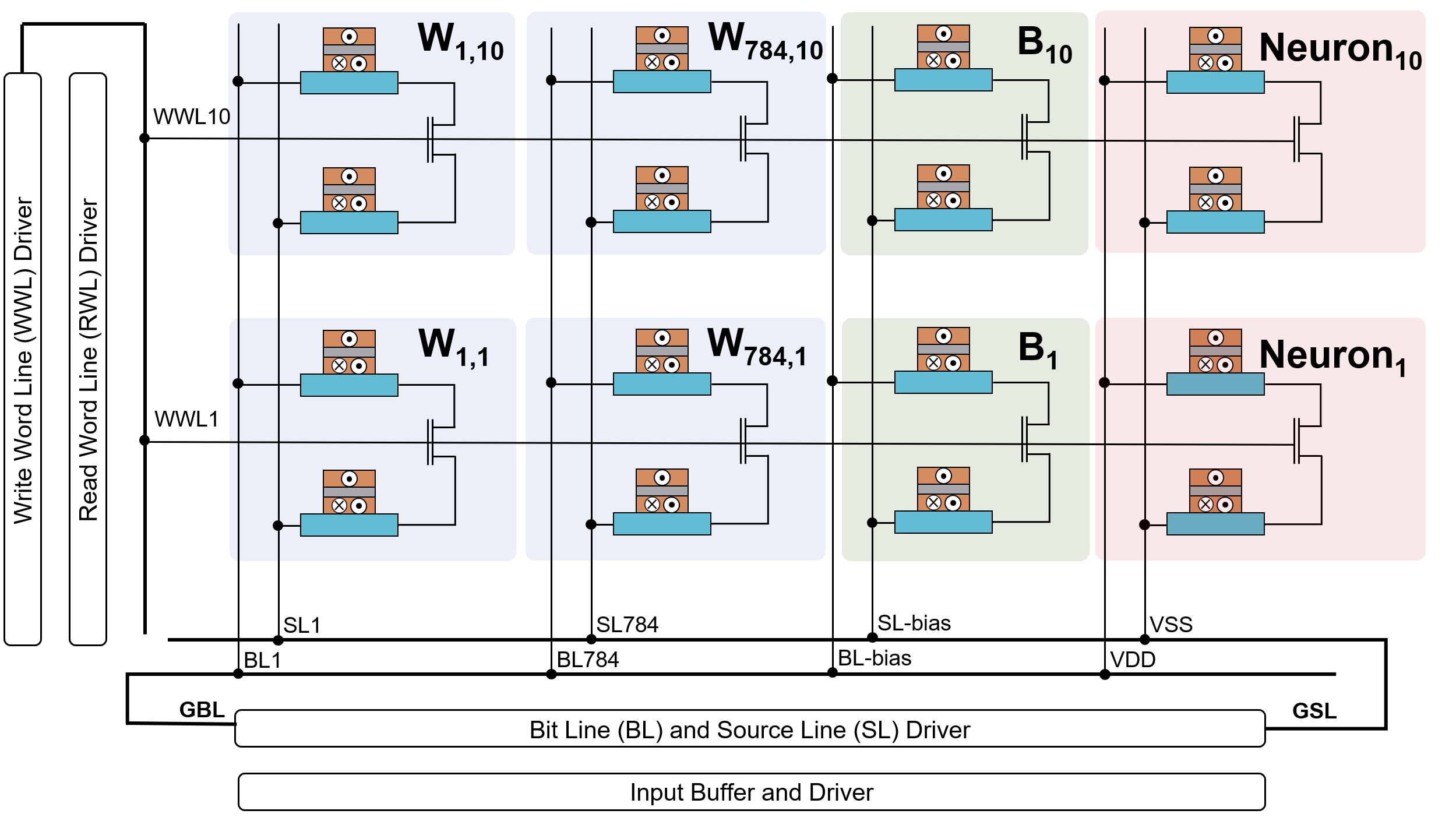}
\vspace{-3mm}
\caption{The training path for a $784 \times 10$ SOT-MRAM based perceptron.}
\label{fig:writearch}
\end{figure}

\begin{figure}[!t]
\centering
\includegraphics[width=3.4in, height=2.4in]{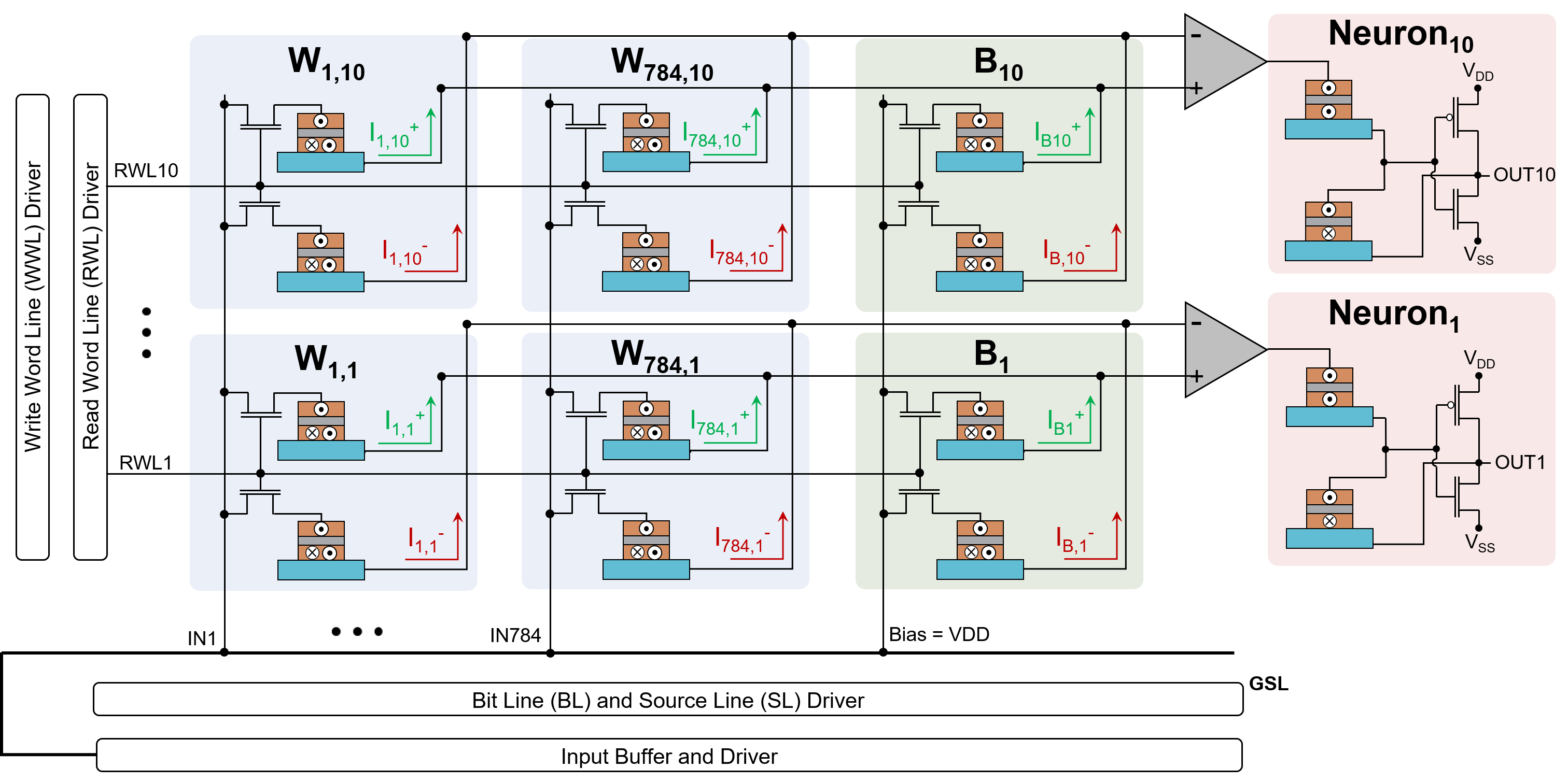}
\vspace{-3mm}
\caption{The inference path for the a $784 \times 10$ SOT-MRAM based perceptron.}
\label{fig:readarch}
\end{figure}

Figures \ref{fig:writearch} and \ref{fig:readarch} exhibit the training and inference paths of a $784 \times 10$ SOT-MRAM based single layer perceptron proposed here, which are shown separately for simplicity. The synaptic connections are designed in form of a crossbar architecture, in which the number of columns and rows are defined based on the number of nodes in input and output layers, respectively. During the training phase, the resistance of the SOT-MRAM based synapses will be tuned using the bit-line (BL) and source-line (SL) interconnections which are shared between different rows, as shown in Fig. \ref{fig:writearch}. The write word line (WWL) control signals will only activate one row in each clock cycle, thus the entire array can be updated using $j$ clock cycles, where $j$ is equal to the number of neurons in the output layer. Moreover, to tune the states of the SOT-MRAMs in neurons according to the requirements mentioned in Section II.A, the BL and SL control signals for the neuron are set to VDD and VSS, respectively, as shown in Fig. \ref{fig:writearch}.

In the inference phase, the BL and SL control signals are in high-impedance (Hi-Z) state, and read word line (RWL) and WWL control signals are connected to VDD and GND, respectively. This will stop the write operation in synapses, and generate $I^+$ and $I^-$ currents shown in Fig. \ref{fig:readarch}, amplitude of which depend on the input (IN) signals and the resistances of SOT-MRAM synapses. Each row includes a shared differential amplifier, which generates an output voltage proportional to $\sum_{i} (I_{i,n}^+-I_{i,n}^-)$ for the $n$th row, where $i$ is the total number of nodes in the input layer. Finally, the output of the differential amplifiers are connected to the SOT-MRAM based sigmoidal neurons. The entire inference operation occurs in parallel and in a single clock cycle. The required signaling to control the training and inference operations is listed in Table \ref{tab:signaling}. One of the main advantages of the proposed architecture is that it can be readily concatenated to form a multi-layer perceptron (MLP) and deep neural network (DNN), which can still operate in a single clock cycle as it will be shown in the Simulation Results section.

\begin{table}[!t]
\caption{The required signaling to control the proposed SOT-MRAM based perceptron array.}
\label{tab:signaling}
\centering
\begin{tabular}{lcccccc}
\hline
\multicolumn{2}{l}{\textbf{Operation}}    & \textbf{WWL} & \textbf{RWL} & \textbf{BL}   & \textbf{SL}   & \textbf{IN}   \\ \hline
\multirow{2}{*}{Training} & $\textbf{W}_i=+1$ & VDD & GND & VDD  & GND  & Hi-Z \\
                          & $\textbf{W}_i=-1$ & VDD & GND & GND  & VDD  & Hi-Z \\ \hline
\multicolumn{2}{l}{Inference}    & GND & VDD & Hi-Z & Hi-Z & VIN  \\ \hline
\end{tabular}
\vspace{-3mm}
\end{table}

\section{Hardware-aware Learning Mechanism for Proposed SOT-MRAM based MLP Architecture}
%As proposed in Section II, we utilize SOT-MRAM devices to realize binarized ($\textbf{W}_i \in \{-1, +1\}$) weights, as well as a sigmoidal activation function. 

To train the proposed SOT-MRAM based neuromorphic MLP architecture, a hardware-aware learning mechanism should be developed which realizes the characteristics and limitations of our SOT-MRAM based neurons and synapses. Herein, we utilize a two stage teacher-student approach, in which both teacher and student networks have identical topologies. Table \ref{tab:learning} provides the notations and descriptions for both teacher and student networks, in which $x$ is the input and $o_i$ is the output of the $i$th neuron.

To incorporate the features of the SOT-MRAM based synapses and neurons within our training mechanism, we have made two modifications to the approaches previously used for training binarized neural networks \cite{xnor-net,binaryconnect}. First, we have used binarized biases in the student networks instead of real-valued biases. %  which could not be implemented by the two SOT-MRAM circuit realization of the biases shown in Fig. \ref{fig:readarch}.
Second, since our SOT-MRAM neuron realizes real-valued sigmoidal activation function ($sigmoid(-x)$) without any computation overheads, we could avoid binarizing the activation functions and reduce the possible information loss in the teacher or student networks \cite{xnor-net}. Herein, after each weight update in the teacher network we clip the real-valued weights within the $[-1,1]$ interval, and then use the below deterministic binarization approach to binarize the weights:

\begin{equation} 
\small
\label{Eq:deter_bin} 
W_{ij} = 
\begin{cases} 
+1, & \bar w_{ij} \ge \Delta_B  \\ 
-1, & \bar w_{ij}< \Delta_B
\end{cases} 
\end{equation}

% \begin{equation} 
% \small
% \label{Eq:deter_ter} 
% W_{ij} = 
% \begin{cases} 
% +1, & \bar w_{ij}>\Delta_T  \\ 
% 0, & |\bar w_{ij}| \le \Delta_T\\
% -1, & \bar w_{ij}<-\Delta_T
% \end{cases} 
% \end{equation}

% \usepackage{multirow}
\begin{table}[]
\caption{The notations and descriptions of the proposed learning mechanism for the SOT-MRAM based MLP.}
\vspace{-2mm}
\label{tab:learning}
\centering
\begin{tabular}{lcc}
\hline
\multirow{2}{*}{}                                             & \multicolumn{1}{c}{\multirow{2}{*}{Teacher Network}} & \multicolumn{1}{c}{\multirow{2}{*}{Student Network}}                                                                      \\
                                                              & \multicolumn{1}{c}{}                                 & \multicolumn{1}{c}{}                                                                                                      \\ \hline
Weights                                                       & $\textbf{W}_i \in R$                                 & $\textbf{W}_i \in \{-1, +1\}$\\ \hline
Biases                                                        & $\textbf{B}_i \in R$                                 & $\textbf{W}_i \in \{-1, +1\}$ \\ \hline
\begin{tabular}[c]{@{}l@{}}Transfer Function\end{tabular}  & $y_i=\textbf{w}_i x + \textbf{b}_i$
& $y_i=\textbf{w}_i x + \textbf{b}_i$                                                                                                                \\ \hline
\begin{tabular}[c]{@{}l@{}}Activation Function\end{tabular} & $o_i=sigmoid(-y_i)$                                      & $o_i=sigmoid(-y_i)$                                                                                                           \\ \hline
\end{tabular}
\end{table}

\noindent where $\Delta_B=0$ is threshold parameters for binarized weights. Finally, once all the binarized weights are trained we will use a mapping mechanism to convert them to resistive states in SOT-MRAM based synapses as explained in Section II.B. Stochastic binarization \cite{binaryconnect} scheme can also be used to quantize the weights and biases. However, stochastic approach exhibits its advantages in very large scale convolutional neural networks (CNNs) which are not the focus of this paper. In fact, we have initially leveraged stochastic mechanisms in our simulations and while the training times were approximately 10-fold longer, the obtained accuracy values were comparable to those realized by deterministic approaches. %Thus, we have used deterministic quantization methods in the simulation results provided in the following section.

\section{Simulation Results}
To evaluate the performance of our proposed SOT-MRAM based neuromorphic MLP architecture, we have utilized a hierarchical simulation approach including circuit-level and application-level simulations as described in the following.

\subsection{Circuit-Level Simulation of SOT-MRAM based Neuron}
Herein, we have used SPICE circuit simulator with 14nm HP-FinFET PTM transistor library, Verilog-A model of the SOT-MRAM using, and $V_{DD}=0.8$ as the nominal voltage to obtain the power consumption of our proposed SOT-MRAM based sigmoid neuron. The obtained simulation results show the average power consumption of $64 \mu W$ for the proposed sigmoid neuron. Moreover, the area of our neuron is approximately equal to $13\lambda \times 30\lambda$, that is obtained by the layout design, in which $\lambda$ is a technology-dependent parameter. Herein, we have used the 14nm FinFET technology, which leads to the approximate area consumption of $0.02 \mu m^2$ per neuron. SOT-MRAM devices can be fabricated on top of the transistors, thus incurring no area overhead

Table \ref{tab:comparison} provides a comparison between our SOT-MRAM based sigmoidal neuron and some of the most power- and area-efficient mixed-signal sigmoid neuron designs. To provide a fair comparison in terms of area and power dissipation, we have utilized General Scaling method \cite{Stillmaker2017Scaling7nm} to normalize the power dissipation and area of the designs listed in Table \ref{tab:comparison}. voltage and area scale at different rate of $U$ and $S$, respectively. Thus, the power dissipation is scaled with respect to $1/U^2$ and area per device is scaled according to $1/S^2$ \cite{Stillmaker2017Scaling7nm}. The results obtained exhibit that the proposed SOT-MRAM based neuron can achieve significant area reduction, while realizing comparable power consumption compared to the existing power- and area-efficient neuron implementations. This results in a $74\times$ and $12\times$ reduction in power-area product compared to the designs introduced in \cite{neuron1} and \cite{neuron2}, respectively.     

\begin{table}[]
\centering
\caption{Performance Comparison for Various Neuron Implementations.}
\label{tab:comparison}
\begin{tabular}{lccc}
\hline
                   & \cite{neuron1} & \cite{neuron2} & Proposed Herein \\ \hline
Power Consumption  & 7.4$\times$      & 0.98$\times$     & 1$\times$                                                        \\
Area Consumption   & 10$\times$       & 12.3$\times$     & 1$\times$                                                        \\ \hline
Power-Area Product & 74$\times$       & 12$\times$       & 1$\times$                                                        \\ \hline
\end{tabular}
\vspace{-3mm}
\end{table}

\subsection{Application-level Simulation}
\begin{figure}[!t]
\centering
\includegraphics[scale=0.34]{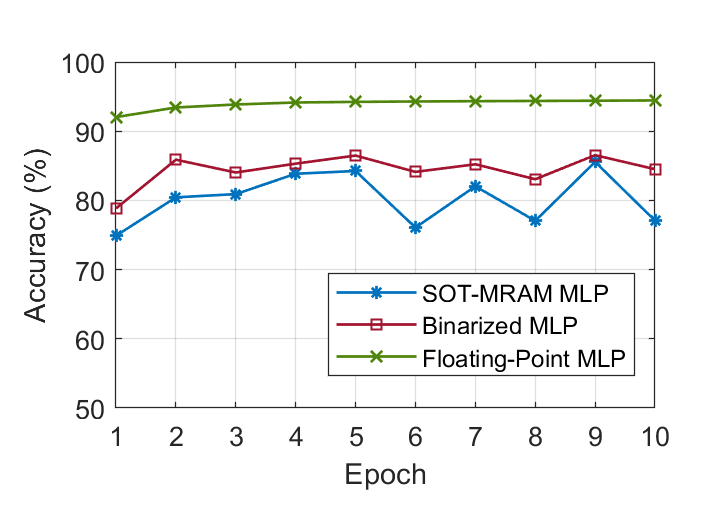}
\vspace{-3mm}
\caption{Accuracy for MNIST application using a $784\times16\times10$ MLP.} \label{fig:acc}
\vspace{-3mm}
\end{figure}

To verify the functionality of our SOT-MRAM based neuron and synapse for larger-scale applications, we have developed a Python-based simulation framework based on \cite{zandjetc2019}. The developed simulator realizes the SPICE circuit implementation of our SOT-MRAM based MLP, and measures its corresponding accuracy and power consumption for a specific pattern recognition application. Fig. \ref{fig:acc} depicts the accuracy of a $784\times16\times10$ SOT-MRAM based neuromorphic MLP simulated in SPICE compared to floating-point and binarized MLP architectures implemented by GPU for MNIST hand-written digit recognition application. The results obtained show that within 10 training epochs a comparable test accuracy of 86.54\% and 85.56\% can be achieved for binarized MLP and SOT-MRAM based MLP architectures, respectively. However, the SOT-MRAM based MLP complete the recognition task in a single clock cycle, while a highly-parallel implementation of binarized MLP on GPU requires $\sim 10^5$ clock cycles with similar frequency to complete the same task.  

%The results obtained validates the functionality and efficiency of our proposed SOT-MRAM based neurons and synapses while being used in larger circuits. 

%Obviously, better accuracies can be achieved by increasing the size of the network which is not the focus of this work.   

\section{Conclusion}
Herein, we proposed a power- and area-efficient SOT-MRAM based sigmoidal neuron, which have been leveraged along-with SOT-MRAM based synapses to construct a neuromorphic MLP architecture. The developed neuron played an enabling role in the single-cycle operation of the SOT-MRAM based MLP. We implemented the SPICE circuit realization of a $784\times16\times10$ SOT-MRAM based MLP and compared its performance with a binarized MLP implemented on GPU for MNIST pattern recognition application. The results obtained exhibited approximately five orders of magnitude increase in the processing speed of our SOT-MRAM based MLP, while realizing comparable accuracy to that of the GPU-implemented binarized MLP. Herein, we have used a small network as a proof-of-concept, while the achieved improvements are expected to be even more significant for larger scale circuits which will be studied in the future work of authors.

\bibliographystyle{IEEEtran}

\balance
\bibliography{ref}
%\begin{thebibliography}{1}


%\end{thebibliography}

\end{document}